# Magneto-optical Kerr effect in Metasurface Flexible Microarrays by near field excitation

N. S. Shnan[1,2], N. Roostaei[1], S. M. Hamidi[1]*, V. I. Belotelov[3,4], A. I. Chernov[5]

[1]Magneto-plasmonic Lab, Laser and Plasma Research Institute, Shahid Beheshti University, Tehran, Iran

[2]Department of Laser Physics, College of Science for Woman, University of Babylon, Babylon, Iraq

[3] Vernadsky Crimean Federal University, 4 Vernadskogo Prospekt, Simferopol, 295007, Russia.

[4] Interdisciplinary Scientific and Educational School of Moscow University «Photonic and Quantum technologies. Digital medicine», Lomonosov Moscow State University, Leninskie gori, 119991 Moscow, Russia

[5] Center for Photonics and 2D Materials, Moscow Institute of Physics and Technology (National Research University), 9 Institutskiy Per., Dolgoprudny 141701, Russia

*E-mail : m_hamidi@sbu.ac.ir

**Abstract**

We have experimentally examined the effect of light localization and near field effect on the magneto-optical response of the two-dimensional coupled micro-ring periodic structure. For this purpose, we fabricated main template by laser writing system and stamped it by polydimethylsiloxane-to reach the main two-dimensional microstructures. Thus, we coated them with a gold layer and Ni layer as plasmonic and magnetic metasurfaces, respectively. We recorded the spectral magneto-optical longitudinal Kerr effect under 200 mT and the spectrometer's response in all visible regions under the normal condition as well as pump and probe system by the aid of green laser as pump. The pump was done via high numerical aperture objective lens to excite near field of plasmon in a two-dimensional structure. Our results indicate that the localized surface plasmon resonance, surface lattice resonance as well as electric and magnetic dipole moments enhance the magneto-optical response in two closer channels in the middle of visible region.

## I. Introduction

Metasurfaces as plasmonic or dielectric have proved to be very practical in holograms [1-3] vortex-phase plates lenses [4-6], tunable functional elements [7-9], and lasers or magneto-photonics [10]. Magneto-optical (MO) properties being useful for isolator or circulators' applications can be tuned or adjusted by external or internal parameters such as magnetic or electric field, plasmonic or dielectric metasurfaces.

So far, the effect of plasmonic metasurfaces (PMS) on the MO response has been investigated in the Au/Co/Au structure by propagative surface plasmon resonance [11], insulator/metal/insulator or plasmonic grating structure [12,13], and surface lattice resonance [14]. In addition, the effect of dielectric metasurfaces (DMS) has been explored on the MO response [ 15,16]

Meanwhile, it is well known that controlling light matter interaction based on the above mentioned plasmonic or dielectric surface waves can be done by designing and constructing different nano or microstructures. These works can be classified into more than four different categories such as use of nanoparticles [17,18], use of multilayer structures [19,20], grating structures [21,22], two dimensional micro arrays [23,24] and so on.

On the other hand, the microarray or coupled micro-ring (CMR) structures instead of usage as wavelength division multiplexing [25], filters [26] and some more telecommunication applications, can be selected as main localization element of light in each motif thus acting as plasmon coupler and increasing the momentum of the incident light to match the plasmonic waves. Depending on the size of the features and spaces between the rings, the absorption, scattering, and diffraction can be controlled which ultimately define light matter interaction and thus control the MO response.

In this paper, we use two-dimensional coupled micro-ring (2D-CMR) structures to excite surface lattice resonance (SLR) of gold perforated structure covered by Ni layer and implemented as a magneto-plasmonic controller.

## II. Experimental Methods

### A. Fabrication Process

A direct laser writing experimental setup utilizing a 405 nm laser beam was employed to fabricate a two-dimensional array of coupled micro-rings (2D-CMR). The laser beam was expanded and illuminated on the silica substrate using an objective lens with a numerical aperture of 0.8. The substrate, coated with SU8 2002 polymer, was placed onto the motorized x-y stage, and the writing process of a two-dimensional micro-rings array onto the polymer was performed. The laser power was set to 0.3 mW to achieve the arrays consisting of 500 in 20 rings with a periodicity of 20 μm, as displayed in Fig. 1(a).

Also, the distance between the two rings and the inner radius of each ring were considered to be 500 nm and 2.70 μm, respectively (Fig. 1(b)). Once writing was completed, a 2D-CMR consisting of coupled micro-rings based on SU8 material arranged on a glass substrate was successfully obtained. To achieve a flexible microarray, the sample fabricated in the previous step was utilized as a stamp, and a flexible microarray based on transparent polydimethylsiloxane (PDMS) polymer was produced via the nanoimprint lithography technique (Fig. 1(c)).

Accordingly, the combination of polydimethylsiloxane and curing agent (with a weight ratio of 10:1) was mixed by a DC mixer for a duration of 5 minutes, and the resulting homogeneous composition was poured onto the micro-rings' array sample. Afterward, the sample was placed in a vacuum chamber for 15 min for degassing.

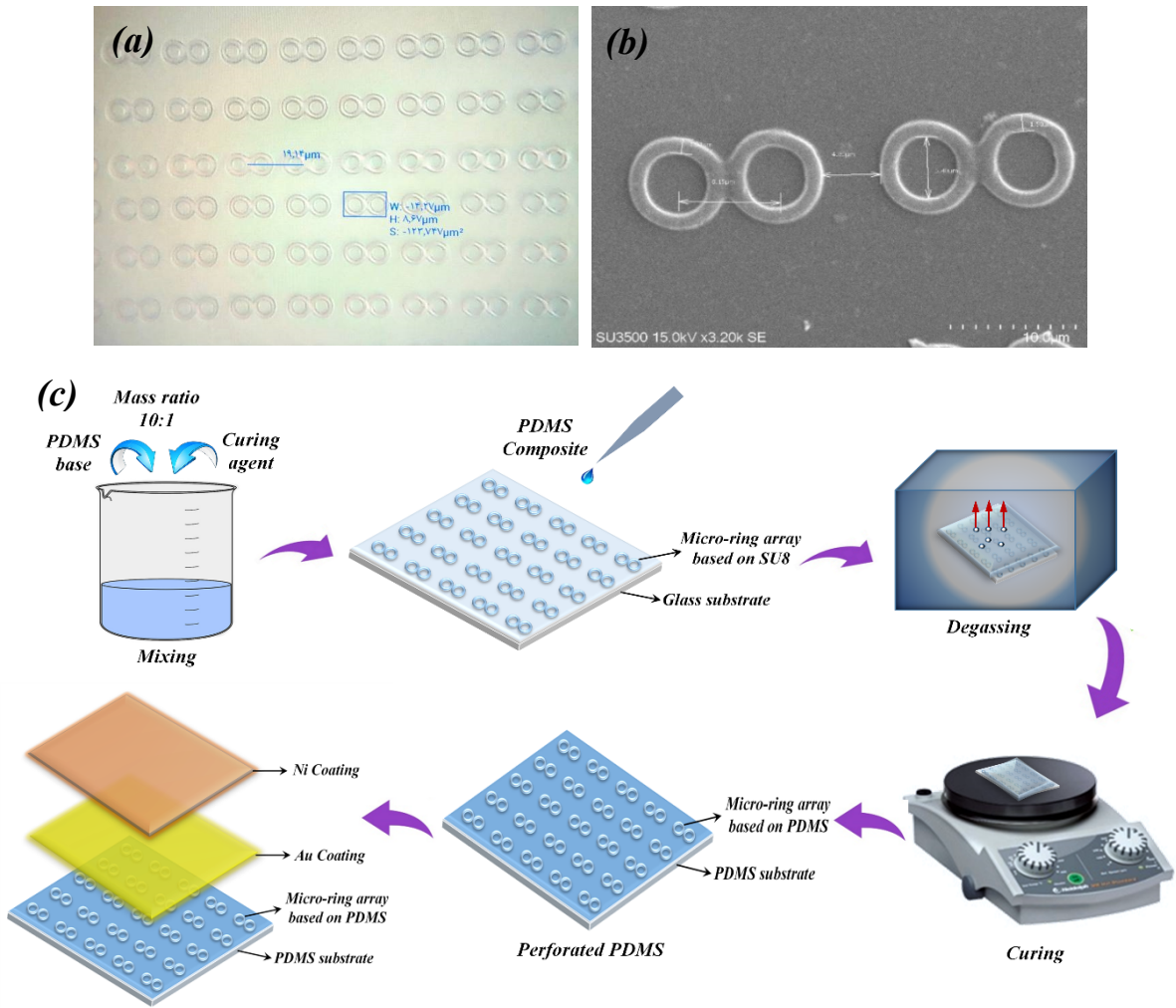

*Figure 1: (a) Image a) dimensions are hard to read. Optical microscopy image of the fabricated micro-ring structure, (b) the SEM image of the proposed micro-rings array, and (c) a schematic array of the soft lithography process for fabrication of the PDMS-based flexible microarray.*

Subsequently, the sample was placed on a hot plate, and the PDMS composite was cured with a gradual increase in temperature ranging from 50 to 100º C over 1 h. After 24 h, the PDMS layer was gently peeled off from the sample, and a flexible PDMS-based microarray was successfully achieved.

Finally, a plasmonic gold thin film with a thickness of 35 nm and a magnetic Ni thin film were deposited onto the flexible microarray via the DC sputtering machine. It should be note that DC sputtering deposition of the thin gold film was carried out under the conditions of a

DC voltage of 365V, plasma current of 0.01mA, chamber pressure of 0.005 mbar, and substrate rotation speed of 28 rpm. Also, DC sputtering deposition of Ni thin film was performed under base chamber pressure of $10^{-5}$ mbar and in Argon gas media, and substrate rotation speed of 28 rpm. It means that we have two main 2D-CMR samples covered with gold as 2D-CMR/Au and 2D-CMR/Au/Ni respectively.

### B. Experimental Setup

A schematic arrangement of the experimental configuration is depicted in Fig. 2. As illustrated in this figure, the polarizer linearly polarizes the un-polarized light, and the polarized light is irradiated onto the sample in the presence of the magnetic field. Afterward, the change in the polarization state is controlled by the analyzer which is collected by the spectrometer. As shown schematically, the reflected light from the sample is coupled to an optical fiber and then recorded using a spectrometer (Ocean Optics) under right and left applied external field in the order of 200 mT.

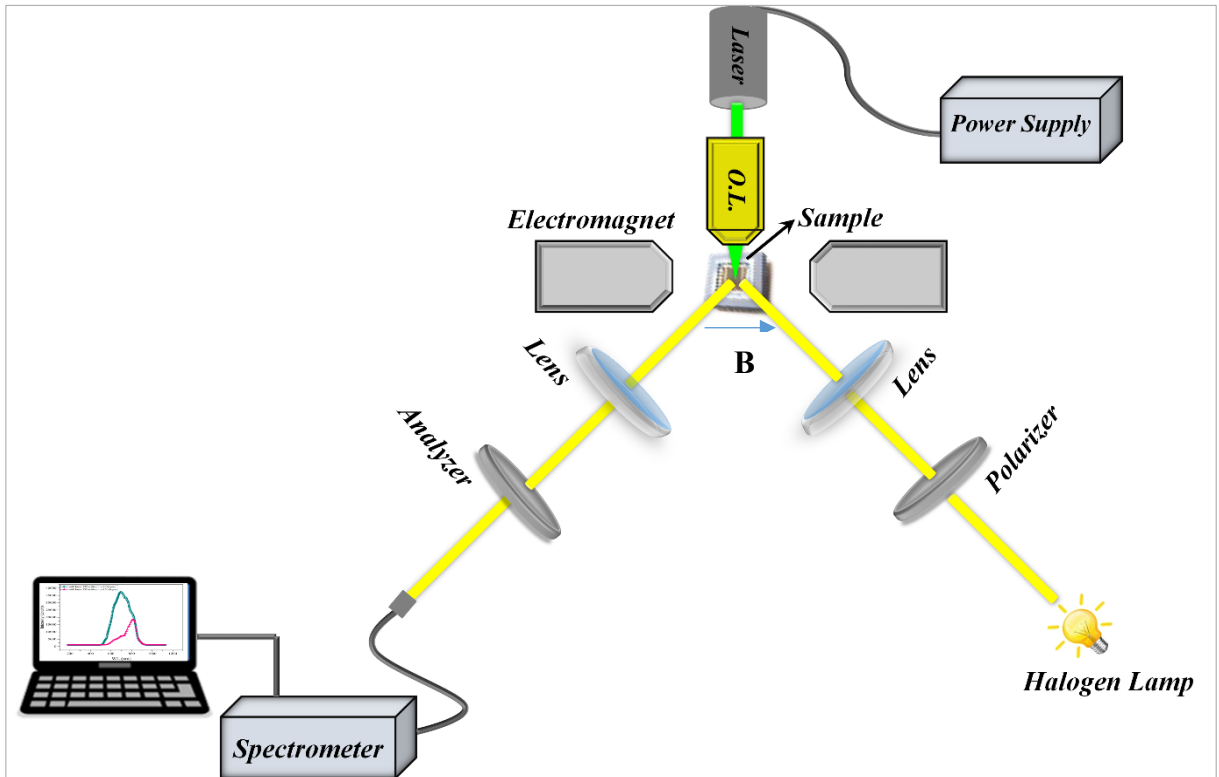
*Figure 2: A schematic array of the Longitudinal magneto-optical Kerr experimental setup.*

All experiments were carried out at room temperature such as our earlier reports paper [10]. Further, a green laser was utilized for optical excitation and enhancement of magneto-optical properties due to thermo-plasmonic effect, which was focused on the sample by an objective lens with a high numerical aperture (NA).

### III. Results and Discussion

As mentioned in the previous section, a 2D-CMR samples of 2D-CMR/Au and 2D-CMR/Au/Ni were placed in the optical and magneto-plasmonic setup to record MO response in the normal state or in the presence of laser pumping via high NA objective lens. By using a high NA excitation, we have a near field excitation of 2D-CMR structure that results in more light localization in each motif of the structure, light matter enhancement and thus improvement of the MO response. This is coming from this fact that by an objective lens with a high numerical aperture (NA = 1.25 in our setup) can be used for enhancement in the spatial resolution over a smaller area.

For the first step, we should record optical response of the sample without any external magnetic field in different analyzer angle, which should have swept from 0 to 180 degree in respect of the first polarizer [12]. As shown in Fig. 3(a) and (b) for two selective angles, we have comparison of optical response in two on and off states of the pumping green laser which confirms reduction in the overall intensity as the default result where some new optical modes appeared by pump green laser. It should be mentioned that we eliminated pump laser signal in the recording process.

The proposed structure consists of regular two-dimensional arrays of pairs of micro-rings, which are coated with a thin layer of gold as a plasmonic material and a thin layer of nickel as a magnetic material. The unit cell consists of a double micro-ring separated by a distance of 400 nm. The experimental reflection spectrum of the fabricated structure is shown in Figures. 3, confirming different modes attributed to the localized surface plasmons resonances (LSPRs) and the coupling effect between the rings. It is worth mentioning that each micro-ring within the structure can support localized surface plasmon resonance (LSPR) modes, and the coupling between the double rings enhances the plasmon resonance. Since the fabricated structure consists of a 2D periodic array of double rings, the coupling between the LSPR induced by the micro-rings will cause the lattice resonance of plasmons. As shown in Figures 3(a), we have the comparison of white light's reflection response in two on and off states of the pumping green laser which confirms the decrease in the overall intensity as by default results in the appearance of new optical modes induced by the pump green laser. In addition, the second derivative of optical signal was used to predict possible wavelengths, assigned by blue circles, with efficient light matter interaction and thus MO response as shown in Figure 3(b).

After that, the spectral MO response of these two samples were recorded as depicted in Fig. 3. In part (c) of this figure, one can see the MO response of 2D-CMR/Au sample as the main reference sample, without any laser pumping, showing fine dispersion-like MO activity from -0.2 to 0.5 degree due to light localization in each motif of the periodic structure of microarrays around 500 nm as main plasmonic wavelength of any gold nanostructure. Then, again the 2D-CMR/Au/Ni thin film was examined under the same condition which repeat the dispersion-like response but at around 580 nm from -0.56 to 2.5 degree with greater broadening as depicted in Fig. 3(d).

Until now, it has been confirmed that our 2D-CMR nanostructure can support light localization and affect the off-diagonal element of epsilon tensor with the aim of enhancing the MO response as it was expected. Meanwhile, near field excitation by pump green light via objective lens should be examined and the MO response of 2D-CMR/Au/Ni sample recorded again. As displayed in Fig. 4, we have 0.6-degree MO Kerr rotation at 571 nm and by dispersion-like manner reach 2.4 degree at 580 nm.

For near field excitation via objective lens and green light pumping, the laser light is focused by the high NA objective lens at the interface between the sample and the metallic nanostructure film. By considering the relative permittivity of gold nanostructure at 532 nm and calculating the ratio of $k_{sp}$ to $k_{light}$, the ring size or NA range can be obtained in which the dispersion matching condition was satisfactory in near field (in our case, NA range in pupil plane is in the order of 1.09–1.095) [27].

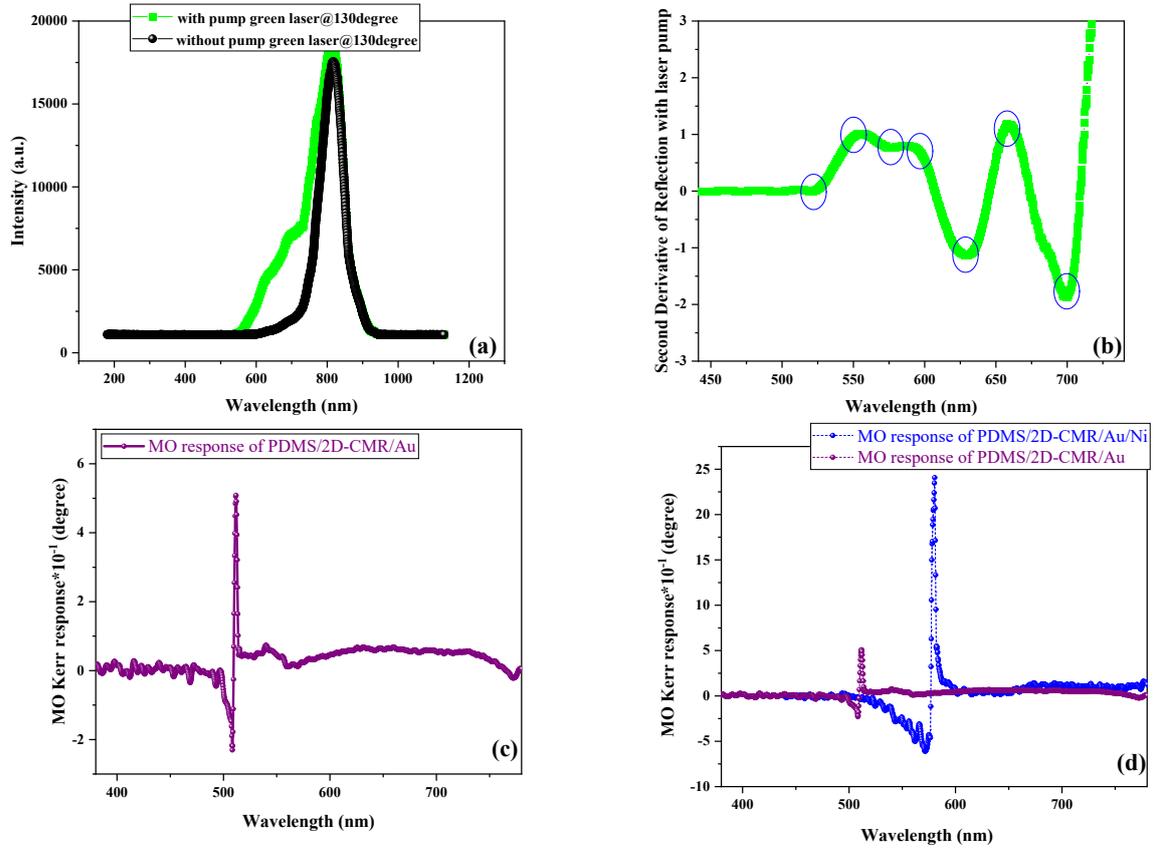

*Figure 3: Optical response of 2D-CMR/Au nanostructure with and without pump green laser and under probe white light at (a) 130 degree difference between two polarizer and analyzer, (b) Derivative of optical response, (c) MO Kerr response of 2D-CMR/Au nanostructure and (d) MO Kerr response of 2D-CMR/Au/Ni nanostructure in the comparison with2D-CMR/Au nanostructure.*

Furthermore, the high NA of the objective satisfies the dispersion matching condition between the light and surface plasmon resonance of 2D-CMR/Au sample can be yield to main shift in the bottom branch of response to 504.9 nm with enhanced Kerr rotation until -4.3 degree and after pass shoulder, reach to 2.8 degree at top branch. It must be note that the main response of the sample without near field excitation, shows again in this figure at 580 nm. Enhancement factor due to near field excitation of 2D-CMR sample is revealed in Figure.4 (b) which confirms again nice enhancement due to thermo-plasmonic effect in the sample in gold nanostructure and light localization in each motif of two dimensional sample.

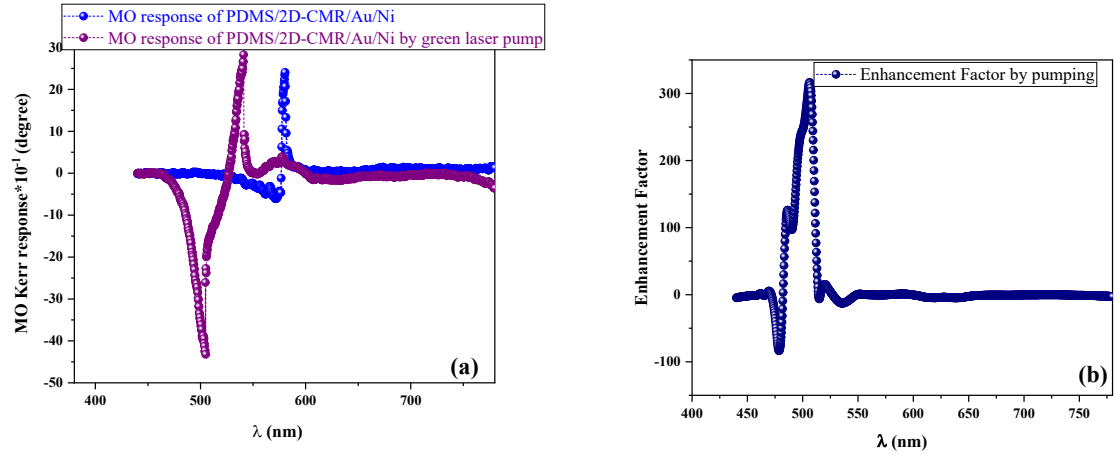

*Figure 4: (a) Magneto-optical response of 2D-CMR/Au/Ni nanostructure under green laser pumping in the comparison with probe only and (b) absolute value of enhancement factor by pumping.*

In proposed plasmonic structure as 2D-CMR/Au, MO response was enhanced due to plasmonic hot spot based on localized surface plasmon resonance (LSPR) in each unit cell and thus surface lattice resonance due to periodic structure. Meanwhile, the electric and magnetic dipoles are essential in the resonance phenomenon in this structure and in 2D-CMR/Au/Ni construction. In the proposed structure, each unit cell of the 2D periodic lattice can support an electric (and magnetic) dipole mode which arise bottom or top resonance response respectively.

## IV. Conclusion

In sum, we fabricated Magneto plasmonic samples based on two-dimensional coupled micro-ring array supported by gold and Ni metasurfaces and investigated the near field excitation by green light pumping. Our results revealed the MO response enhancement in distinct channels based on LSPR of each motif, surface lattice resonance as well as electric and magnetic dipole moments in the middle part of the visible region. This investigation can help us design and fabricate a new generation of MO medium useful for MO isolators and other applications by pump and probe measurement system.

**Conflict of interest:**

There is no any conflict of interest.

## Acknowledgment:

This work has been supported by the center for International Scientific Studies & Collaborations (CISSC) Ministry of science Research and Technology of Iran by number 4020199. In addition, the work was supported by the Ministry of Science and Higher Education (FSMG-2021-0005)